\documentclass[11pt,a4paper]{article}
\usepackage{naacl2021}
\usepackage{amsmath, amssymb, amsthm, amscd, amsfonts}
\usepackage{times}
\usepackage{latexsym}
\usepackage{graphicx}
\usepackage{epstopdf}
\usepackage{algorithm}
\usepackage{algorithmicx}
\usepackage{algpseudocode}
\usepackage{url}
\usepackage{subfigure}
\usepackage{url}
\usepackage{enumerate}
\usepackage{multirow}
\usepackage{bm}
\usepackage{array}
\usepackage[OT1]{fontenc}
\usepackage{bbding}
\usepackage{listings}
\usepackage{xspace}
\usepackage{url}

\usepackage{booktabs}
\usepackage{enumitem}

\usepackage{microtype}

\usepackage{tablefootnote}

\graphicspath{{pics/}}

\newcommand{\toprightarrow}[1]{\mathord{\buildrel{\lower3pt\hbox{$\scriptscriptstyle\rightarrow$}}\over#1} }
\newcommand{\topleftarrow}[1]{\mathord{\buildrel{\lower3pt\hbox{$\scriptscriptstyle\leftarrow$}}\over#1} }

\newcommand{\method}{LightSeq\xspace}
\newcommand{\fastertransformer}{FasterTransformer\xspace}

\title{\method: A High Performance Inference Library for Transformers}

\author{Xiaohui Wang, Ying Xiong, Yang Wei, Mingxuan Wang, Lei Li \\ 
ByteDance AI Lab \\ 
\texttt{\{wangxiaohui.neo, xiongying.taka, weiyang.god\}@bytedance.com} \\
\texttt{\{wangmingxuan.89, lileilab\}@bytedance.com}}

\date{}

\begin{document}
    \maketitle
    \begin{abstract}
        
Transformer, BERT and their variants have achieved great success in natural language processing. 
Since Transformer models are huge in size, 
serving these models is a challenge for real industrial applications.
In this paper, we propose \method, a highly efficient inference library
for models in the Transformer family. 
\method includes a series of GPU optimization techniques to to streamline the computation of neural layers and to reduce memory footprint. 
\method can easily import models trained using PyTorch and Tensorflow. 
Experimental results on machine translation benchmarks show that
\method  achieves up to 14x speedup compared with TensorFlow and 1.4x compared with \fastertransformer, a concurrent CUDA implementation. 
The code is available at \url{https://github.com/bytedance/lightseq}.

    \end{abstract}
    
    \section{Introduction}
    \label{sec:intro}

    \begin{table*}[!ht]
        \normalsize
        \begin{center}
            \resizebox{\linewidth}{!}{
                \begin{tabular}{l|ccccc|cccc}
                    \toprule
                                                                   & \multicolumn{5}{c|}{\textbf{Models}}                                           & \multicolumn{3}{c}{\textbf{Decoding Methods}}                                 \\ 
                    \multirow{-2}{*}{\textbf{Inference Libraries}} & Transformer              & GPT                      & VAE          & BERT & Multilingual            & Beam Search              & Diverse Beam Search      & Sampling                 \\ \midrule
                    FasterTransformer                             & { \Checkmark} & { \Checkmark} & { \XSolidBrush}& { \Checkmark} & { \XSolidBrush} & { \Checkmark} & { \Checkmark} & { \Checkmark} \\ 
                    TurboTransformers                             & { \Checkmark} & { \XSolidBrush} & { \XSolidBrush}& { \Checkmark} & { \XSolidBrush} & { \XSolidBrush } & { \XSolidBrush} & { \XSolidBrush} \\ 
                    \method                               & { \Checkmark} & { \Checkmark} & { \Checkmark}  & { \Checkmark} & { \Checkmark}& { \Checkmark} & { \Checkmark} & { \Checkmark} \\ 
                    \bottomrule
                    \end{tabular}
            }
        \end{center}
        
        \caption{Features for FasterTransformer, TurboTransformers and our proposed \method.
        \method  supports the most features for a comprehensive set of Transformer models.}
        \label{Tab:FeaComp}
    \end{table*}
    
    Sequence processing and generation have been fundamental capabilities for many natural language processing tasks,
    including machine translation, summarization, language modeling, etc
    \citep{DBLP:conf/emnlp/LuongPM15, DBLP:journals/corr/abs-2001-04063, DBLP:conf/acl/DaiYYCLS19}.
    In recent years, with the introduction of Transformer model \citep{DBLP:conf/nips/VaswaniSPUJGKP17},
    many pre-trained language models such as BERT, GPT, and mRASP have also been widely used in these tasks
    \citep{DBLP:journals/corr/abs-1810-04805, radford2019language,yang2020towards,lin2020pre}. 
    
    However,  the parameters  of these models become increasingly large,
    which causes the high latency of inference and
    brings great challenges to the  deployment \citep{DBLP:journals/corr/abs-2010-13382}.
    The current popular inference systems are not necessarily the best choice for the online service of sequence processing problems. 
    First, training frameworks, such as TensorFlow and PyTorch,  require accommodating flexible model architectures and backward propagation, which introduce additional memory allocation and extra overhead of using fine-grain kernel functions. Therefore, the direct deployment of the training framework is not able to make full use of the hardware resource.    
    Taking an example of machine translation, the Transformer big model currently takes roughly 2 seconds to translate a sentence,
    which is unacceptable in both academia and industry
    \citep{DBLP:conf/emnlp/EdunovOAG18, DBLP:journals/corr/abs-2010-02416}.
    Second, current optimizing compilers for deep learning such as TensorFlow  XLA~\cite{abadi2017computational}, TVM~\cite{chen2018tvm} and Tensor RT~\cite{vanholder2016efficient} are mainly  designed for fixed-size inputs. However, most NLP problems enjoy variable-length inputs, which are much more complex  and require dynamic memory allocation. 
    Therefore,  a high-performance sequence inference library for variable-length inputs is required.
    There are several concurrent CUDA libraries which share a similar idea with our project,
    such as FasterTransformer~\footnote{\url{https://github.com/NVIDIA/FasterTransformer}} and
    TurboTransformers ~\citep{DBLP:journals/corr/abs-2010-05680}.
    
We will highlight three innovative features that make
\method outperforms  similar projects.  
First,  we replace a straightforward combination of fine-grained GPU kernel functions in TensorFlow or PyTorch implementations with coarse-grain fused ones, which avoid high time cost introduced by a mass of kernel function launches and GPU memory I/O for intermediate results. 
As a result, \method reduces the atomic kernel functions by four times compared with Tensorflow approaches. 
Second, we specially design a hierarchical auto regressive search method to speed up the auto-regressive search. 
Third, we propose a dynamic GPU memory reuse strategy.  
Different from fixed-length inputs, sequence processing tackles the variable-length inputs, which bring difficulty for memory allocation.  \method proposes to pre-define the maximal memory for each kernel function and shares the GPU memory across non-dependent ones.  
As a result, \method reduces eight times memory allocation  without loss of inference speed. 
As a benefit, \method enjoys several advantages:

\begin{description}
  \item[Efficient] \method  shows better inference performance for generation tasks.
        For example, in machine translation benchmarks, \method  achieves up to 14 times speedup compared with TensorFlow
        and 1.4 times speedup compared with FasterTransformer.  
        
        \item[Functional] \method  supports  more architecture  variants,
        such as BERT, GPT, Transformer, and Variational Autoencoders (VAEs).
        Further, \method provides different search algorithms,
        such as beam search, diverse beam search and probabilistic sampling \citep{DBLP:journals/corr/VijayakumarCSSL16}.
        Table \ref{Tab:FeaComp} shows the functional comparison between FasterTransformer\footnote{As of this writing, we use FasterTransformer v2.1 for comparison.}, TurboTransformers\footnote{we use TurboTransformers for comparison at commit 0eae02ebadc8b816cd9bb71f8955a7e620861cd8},
        and  \method  in text generation tasks.
        \item[Convenient] \method  is easy to use, which contains a serving  system and efficient CUDA implementations.  
        The popular models, such as BERT, Roberta, GPT, VAEs, MT Transformer, and Speech Transformer  can be directly deployed online without code modification. 
        For  user-specific architectures,  \method supports multiple model reuse, which can be easily adapted with only a few lines of code modification. 
\end{description}




    
    \section{\method Approach}
    \label{sec:approach}

    \begin{figure*}[t]
        \centering
        \includegraphics[width=0.7\linewidth]{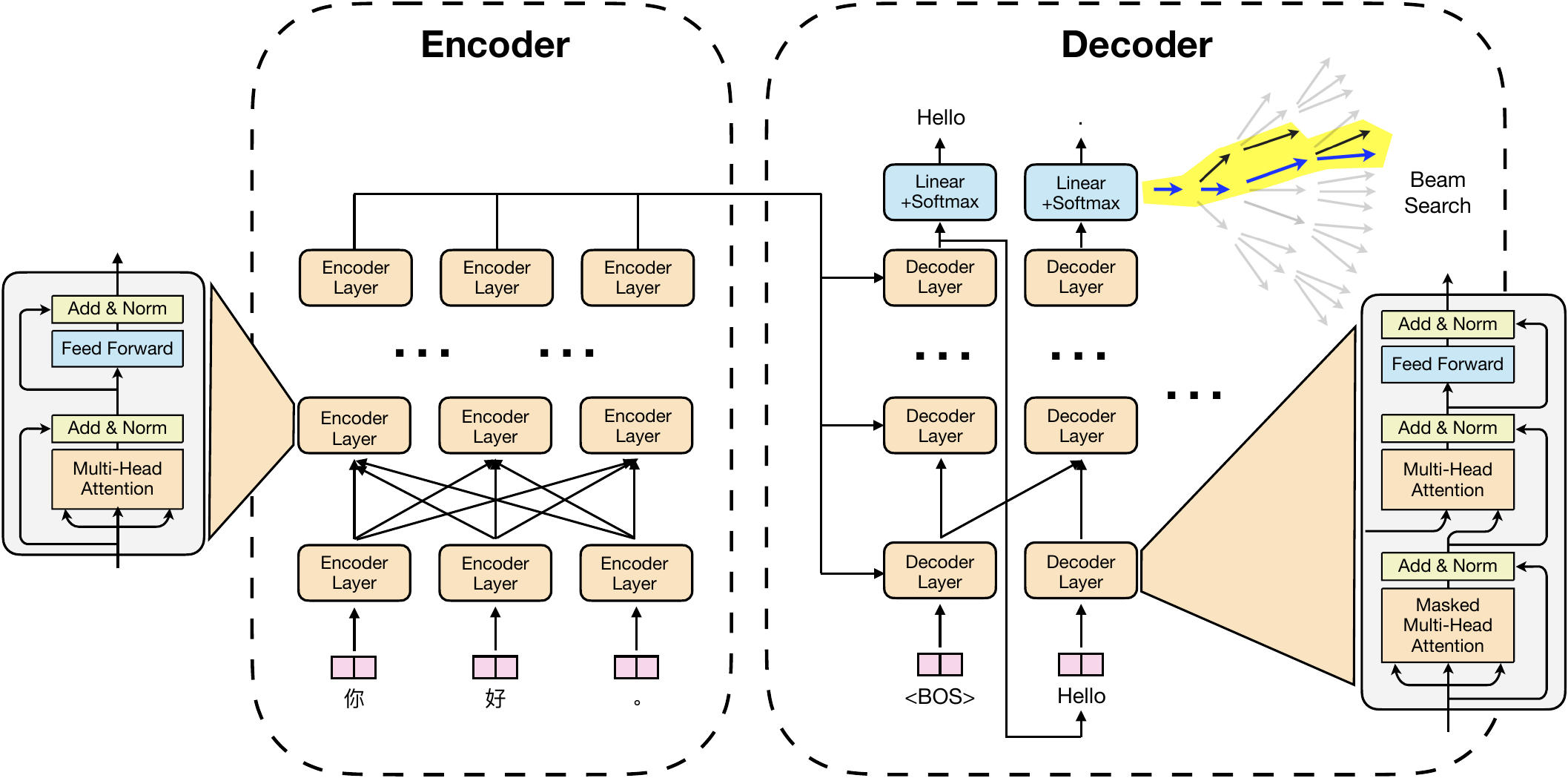}
        \caption{The process of sequence to sequence generation using Transformer model with beam search.}
        \label{Fig:transformer}
    \end{figure*}
    
    Transformer-based NLP models mainly consist of two components during inference: the feature calculation layer and the output layer, as shown in Figure \ref{Fig:transformer}.
    
    The feature calculation layer is mainly based on self-attention mechanism and feature transformation, which is actually implemented by matrix multiplication and a series of I/O-intensive operations such as element-wise (e.g., reshape)
    and reduce (e.g., layer normalization).
    
    The output layer slightly changes in different tasks, such as classification in NLU tasks or search (e.g., beam search) in NLG tasks. This layer is usually composed of the \texttt{Softmax} over vocabulary, probability sorting, cache refreshing, etc., which are essentially I/O-intensive.
    
    These two components pose challenges for efficient inference:
    
    \begin{itemize}

        \item The fine-grained call of I/O-intensive GPU kernel function brings a huge amount of GPU memory I/O, which becomes the performance bottleneck of feature calculation.
        \item Redundant calculations exist due to the fact that we only need a few tokens/labels with the highest probability instead of all in classification or search for the output layer.
        \item Dynamic shape in variable sequence length and auto-regressive search makes it difficult to achieve memory reuse within or between requests, which leads to a large number of GPU memory allocation during model service.
    \end{itemize}

    \method employs a series of innovative methods to address these challenges to accelerate model development,
    such as fusion of multiple kernel functions to reduce I/O overhead, hierarchical optimization of search algorithms to erase redundant calculations, and reuse of dynamic GPU memory to avoid run-time allocation.
    The following is a detailed introduction to these methods.
    
    \subsection{Operation Fusion}
    Transformer feature calculation layer needs to be highly optimized since it is ubiquitous in various NLP tasks today. In most deep learning frameworks, such as TensorFlow and PyTorch, it is implemented by a straightforward combination of fine-grained kernel functions from standard libraries provided by hardware manufacturers, which introduces high time cost due to a mass of kernel function launches and GPU memory I/O for intermediate results.
    
    Taking layer normalization implemented by TensorFlow as an example, there are still three kernel launches\footnote{Two for \texttt{reduce\_mean} operations and one for calculation of the final result.}
    and two intermediate results (mean and variance) even with the help of optimizing compilers like TensorFlow  XLA~\cite{abadi2017computational}.  As a comparison, we can write a custom kernel function dedicated to layer normalization based on the CUDA toolkit, which produces only one kernel launch without intermediate results.

\begin{figure}[t]
        \centering
        \includegraphics[width=0.7\linewidth]{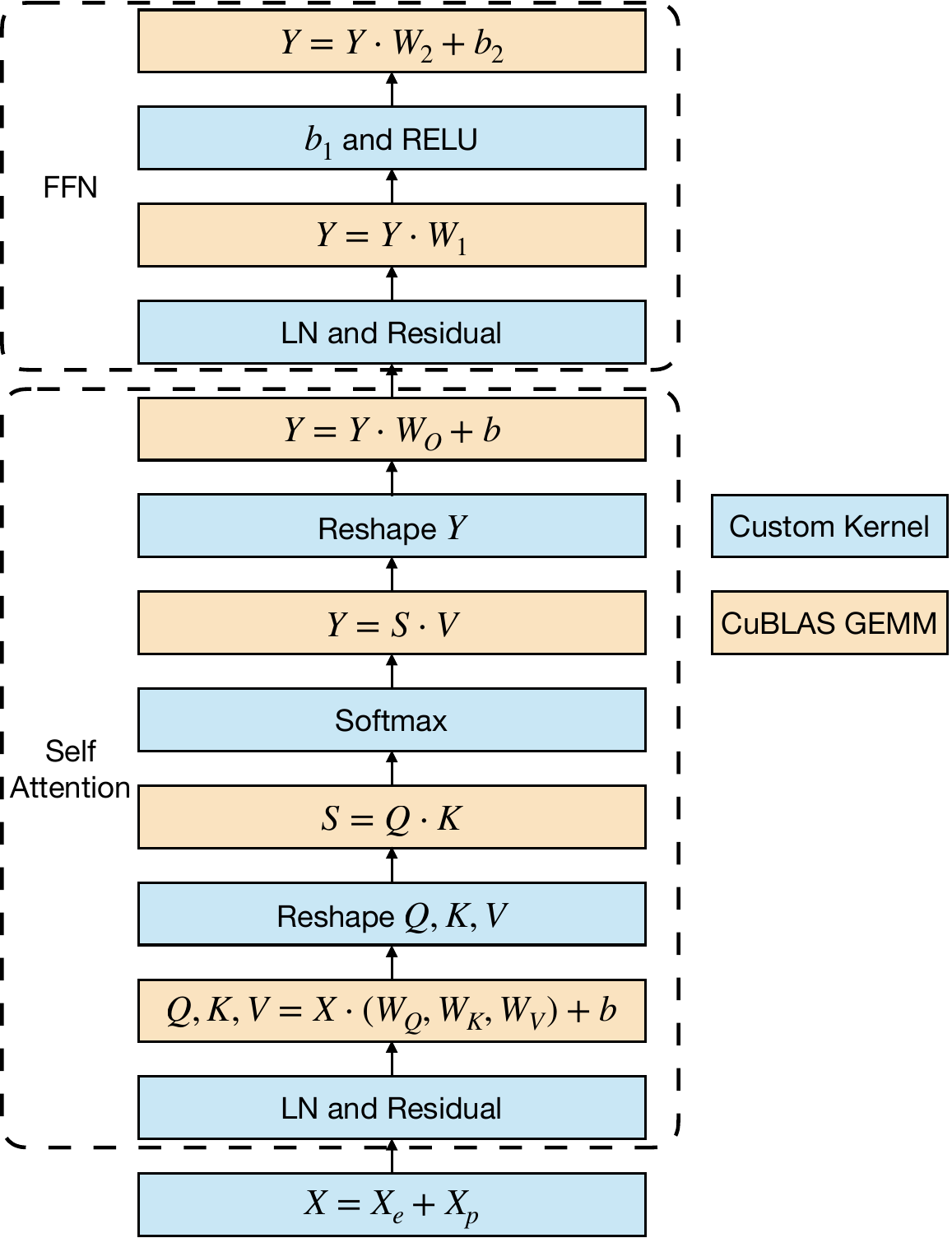}
        \caption{The structure of optimized Transformer encoder layers in \method .}
        \label{Fig:GEMM}
    \end{figure}

    \method implements the Transformer feature calculation layer with general matrix multiply (GEMM) provided by cuBLAS\footnote{\url{https://developer.nvidia.com/cublas}} and custom kernel functions. The detailed structure is shown in Figure \ref{Fig:GEMM}.
    Combination of fine-grained operations between GEMM operations is fused into one custom kernel function. In consequence, there are only six custom kernel functions and six GEMM in a Transformer encoder layer, which is usually more than four times less than its corresponding implementation in common deep learning frameworks like TensorFlow or PyTorch.

    \subsection{Hierarchical Auto Regressive Search}
    
    \method supports a comprehensive set of output layers, such as sentence-level and token-level classification, perplexity calculation for language models, and auto-regressive search like beam search, diverse beam search and top-\(k\)/top-\(p\) sampling \citep{DBLP:journals/corr/abs-1904-09751}. 
    Redundant calculations often exist in these output layers since we only need a few labels/tokens with the highest probability instead of all of them.
    Auto-regressive search is relatively complicated, and we will discuss it in the next paragraph. For the other types of output layers, we can simply replace \texttt{Softmax} with the probability calculation of token/label with the highest \texttt{logits}, which brings more obvious benefit when the size of vocabulary or labels is large.
    
    Auto-regressive search is widely used in machine translation and text generation. \method proposes Hierarchical Auto Regressive Search (HARS) method to erase redundant calculations and parallel computing.
    Here we take the most used beam search method as an example
    to introduce the proposed HARS method.

    In one step of the beam search process, given the \texttt{logits}, we need to perform two calculations over the whole vocabulary:
    
        \begin{enumerate}
        \item Compute the conditional probability using \texttt{Softmax} and write the intermediate result into GPU memory.
        \item Read the intermediate result from GPU memory and select the top-\(k\) beams and tokens by sequential probability.
        \end{enumerate}
 
    These two calculations are highly time-consuming since the vocabulary size is usually in tens of thousands of scales. For example, they account for a latency proportion of 30\% in Transformer base models.
    
    In order to reduce the input size of these two calculations, \method introduces a two-stage strategy that is widely employed in the recommended system: retrieve and re-rank.
    
    Before the probability computation and top-\(k\) selection, the retrieve is carried out first. For each beam, we calculate as follows:
        \begin{enumerate}
        \item Randomly divide \texttt{logits} into $k$ groups.
        \item Calculate the maximum of group $i$, denoted as $m_{i}$
        \item Calculate the minimum of $m_{i}$, denoted as $\mathcal{R}$, which can be regarded as a rough top-$k$ value of \texttt{logits}.
        \item Select \texttt{logits} larger than $\mathcal{R}$ and write them into GPU memory.
        \end{enumerate}
    The retrieve is co-designed based on GPU characteristics and \texttt{logits} distribution. Hence it is efficient and effective:
        \begin{itemize}
        \item Efficient. The retrieve is implemented by one kernel function and can be executed within a dozen instruction cycles.
        \item Effective. After the retrieve, only dozens of candidates were selected.
        \end{itemize}
    After the retrieve, the original two calculations of beam search will be carried out on the small set of candidates, named as Hierarchical Auto Regressive Search.

\begin{figure}[t]
        \centering
        \includegraphics[width=0.7\linewidth]{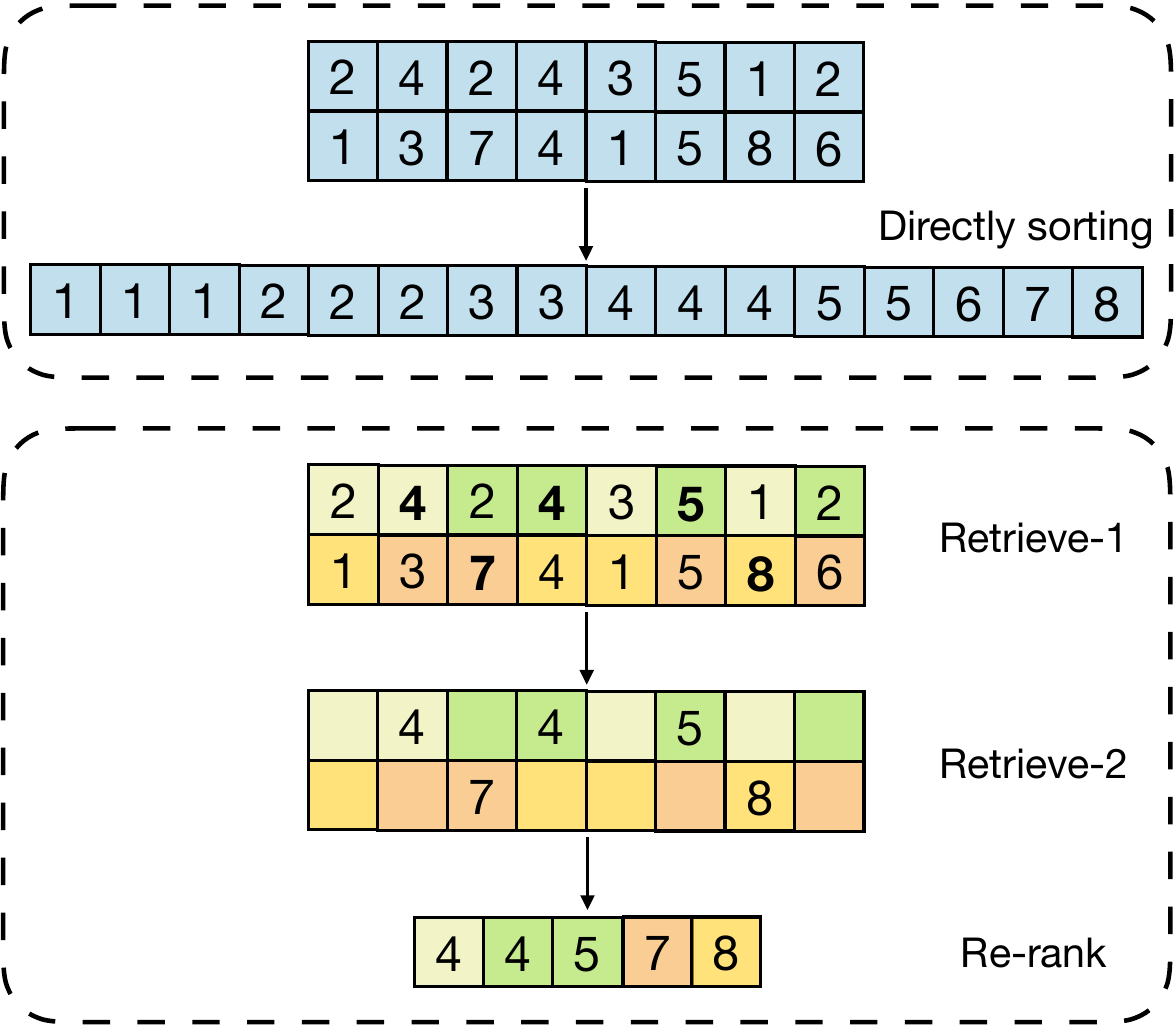}
        \caption{An illustration of the proposed hierarchical strategy. In this case, beam size is 2 and vocabulary size is 8. Each row represents \texttt{logits} in a beam.}
        \label{Fig:softmax}
    \end{figure}

    Figure \ref{Fig:softmax} is a detailed illustration of the proposed hierarchical strategy.
    In the original beam search method, we need to compute the probability and select the top-\(k\) over the whole vocabulary. 
    However, by hierarchical method, we only need to pick a small set of candidates from each beam and then perform probability computation and top-\(k\) selection.
    
    \begin{figure*}[t]
        \centering
        \subfigure[TensorFlow with Float16.]{
            \begin{minipage}[t]{0.27\textwidth}
                \includegraphics[width=\textwidth]{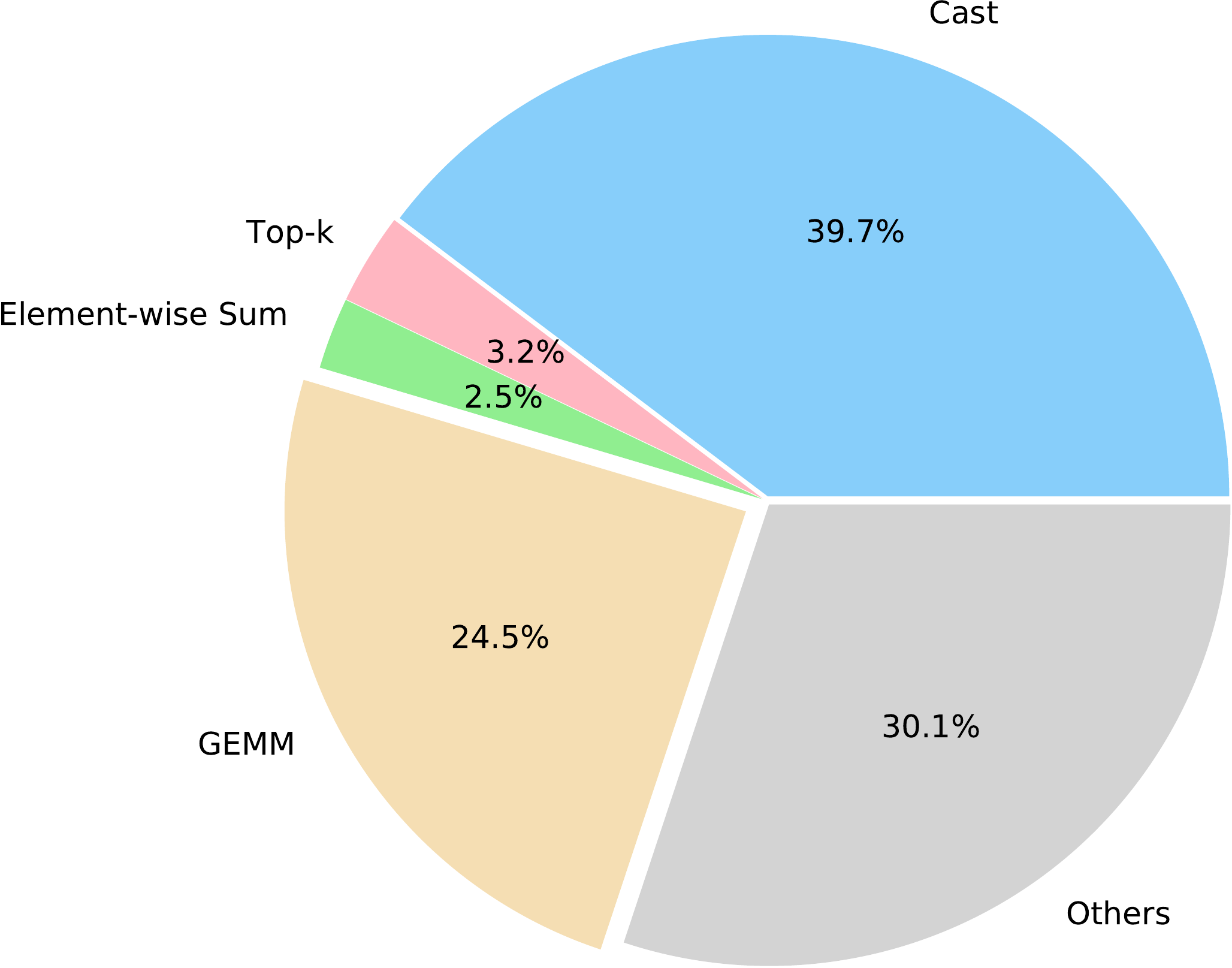}
            \end{minipage}
            \label{Fig:time_1}
        }
        \subfigure[\method with Float16.]{
            \begin{minipage}[t]{0.31\textwidth}
                \includegraphics[width=\textwidth]{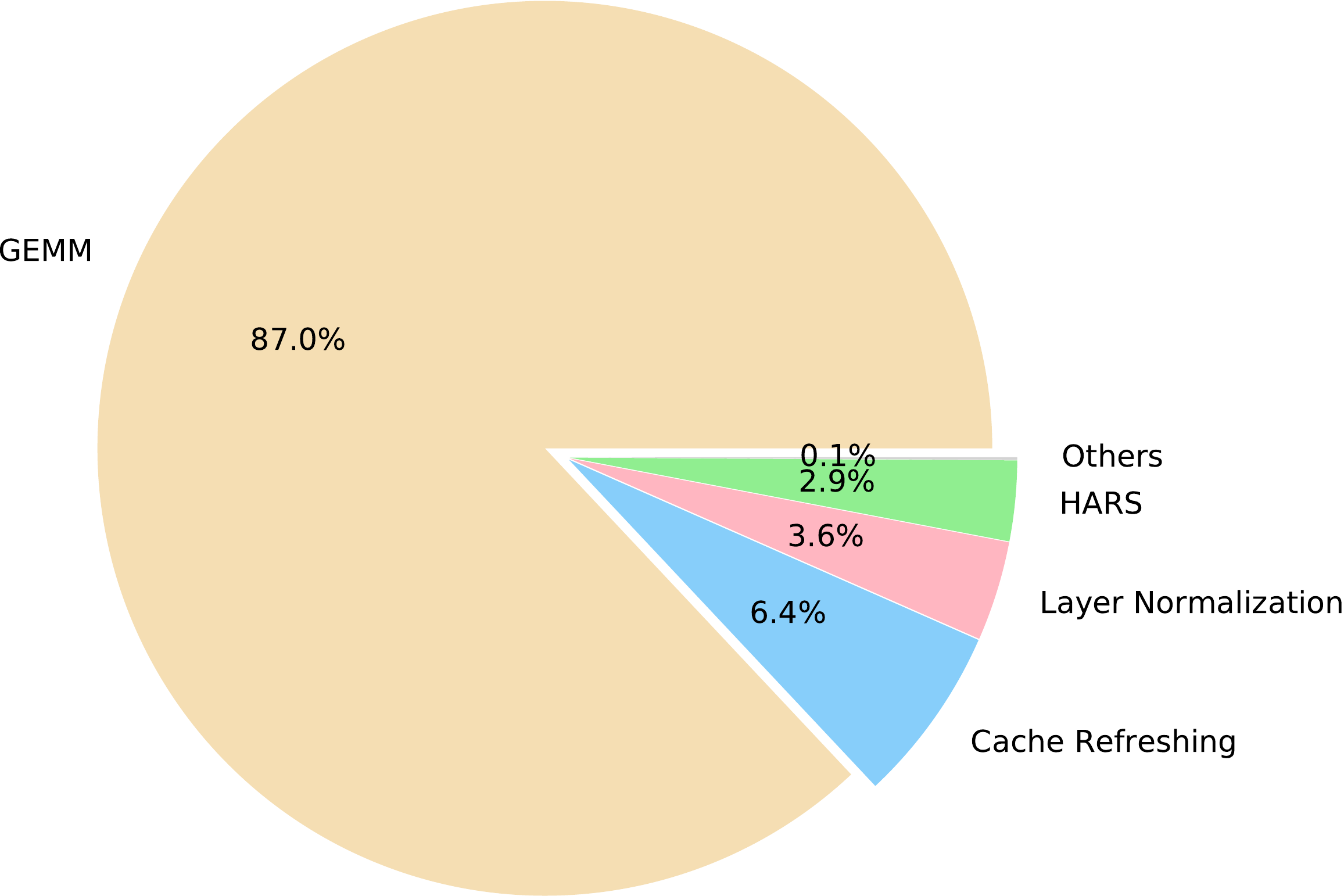}
            \end{minipage}
            \label{Fig:time_2}
        }
        \subfigure[\method with Float32.]{
            \begin{minipage}[t]{0.3\textwidth}
                \includegraphics[width=\textwidth]{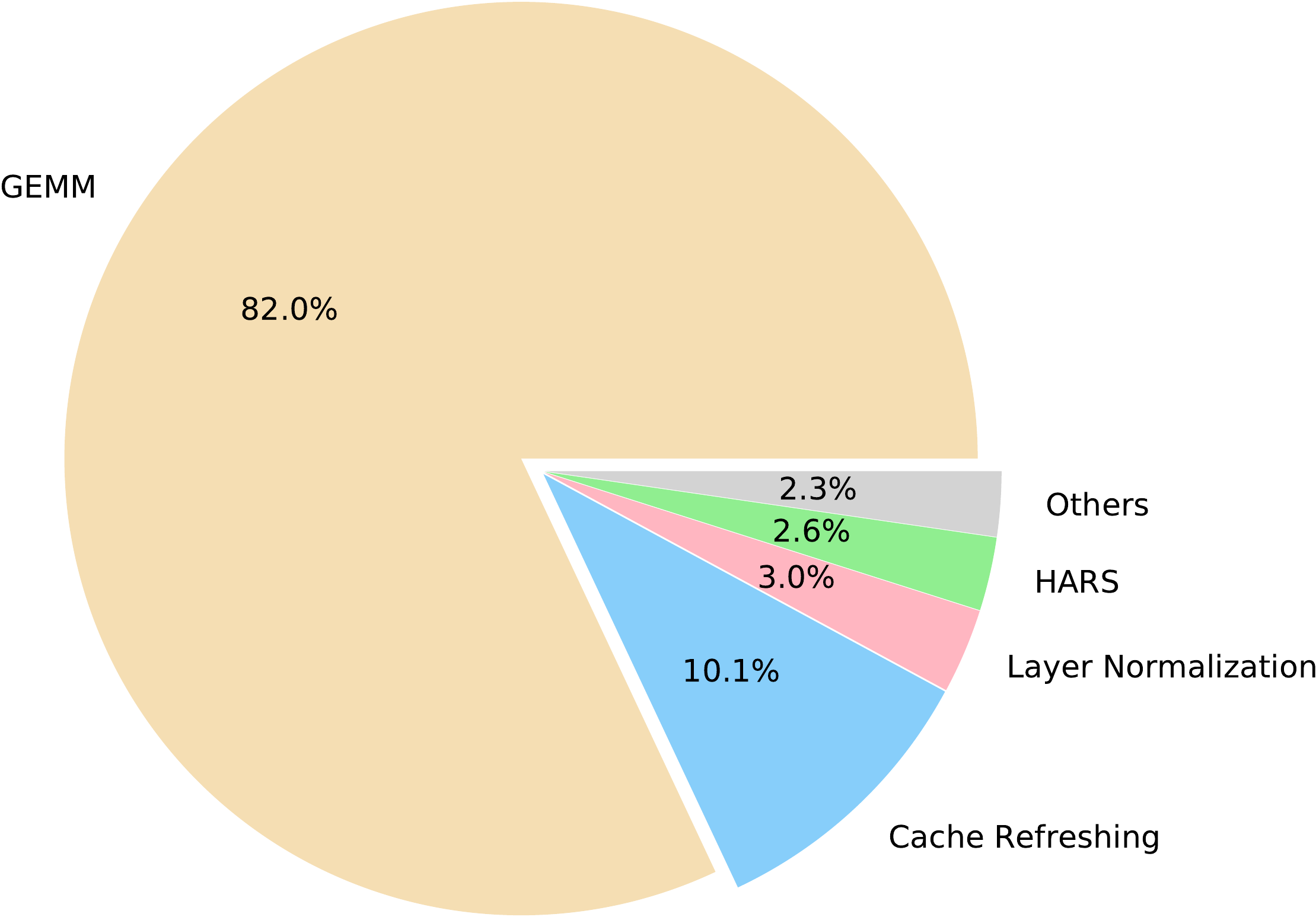}
            \end{minipage}
            \label{Fig:time_3}
        }
        \caption{Proportion of computation occupation. GEMM is the main indicator and the larger number indicates the higher computation efficiency.}
        \label{Fig:model}
    \end{figure*}
    
    \begin{figure*}[t]
        \centering
        \subfigure[P4 speedup in Float32.]{
            \begin{minipage}[t]{0.475\textwidth}
                \includegraphics[width=\textwidth]{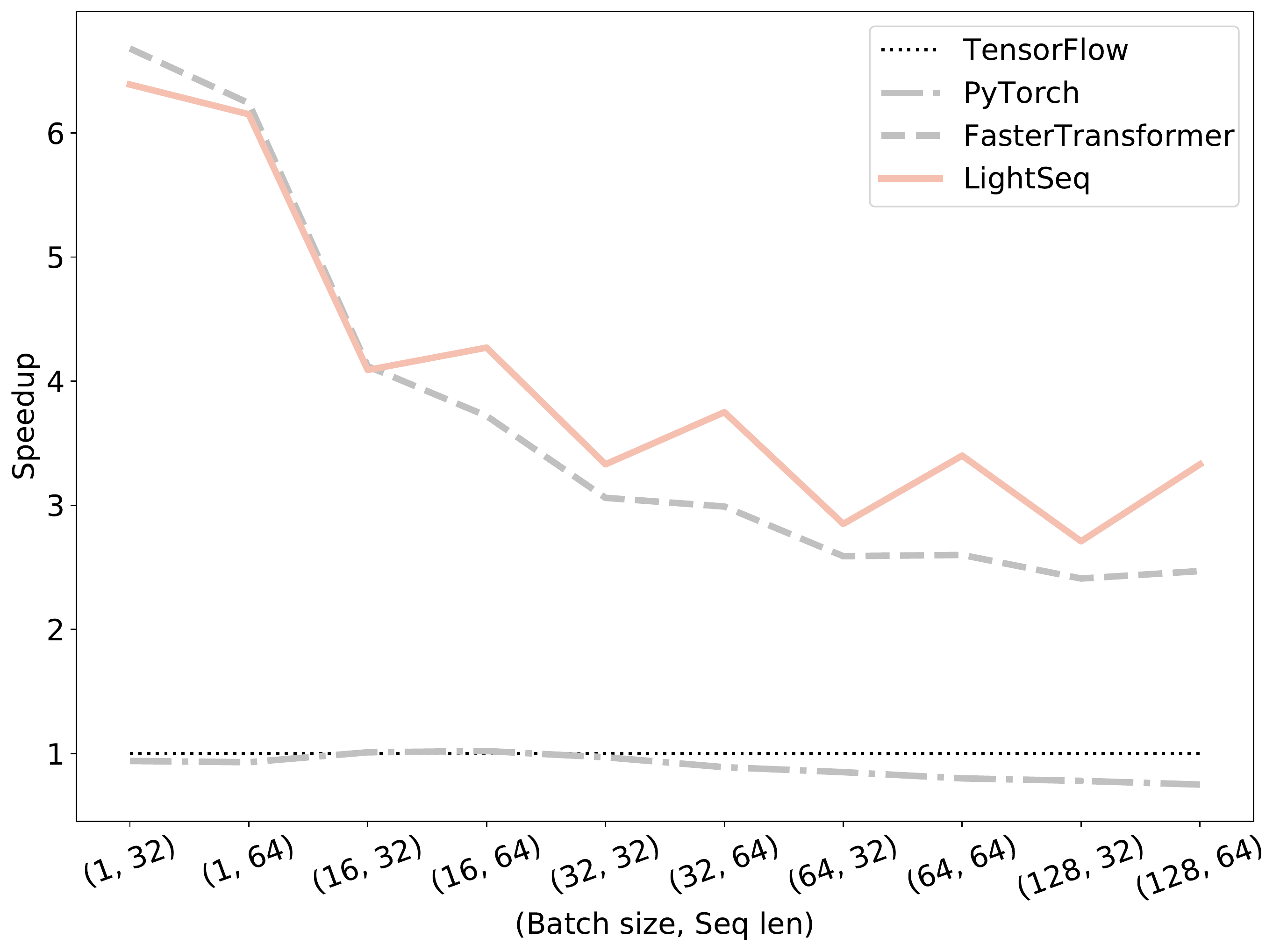}
            \end{minipage}
        }
        \subfigure[T4 speedup in Float16.]{
            \begin{minipage}[t]{0.482\textwidth}
                \includegraphics[width=\textwidth]{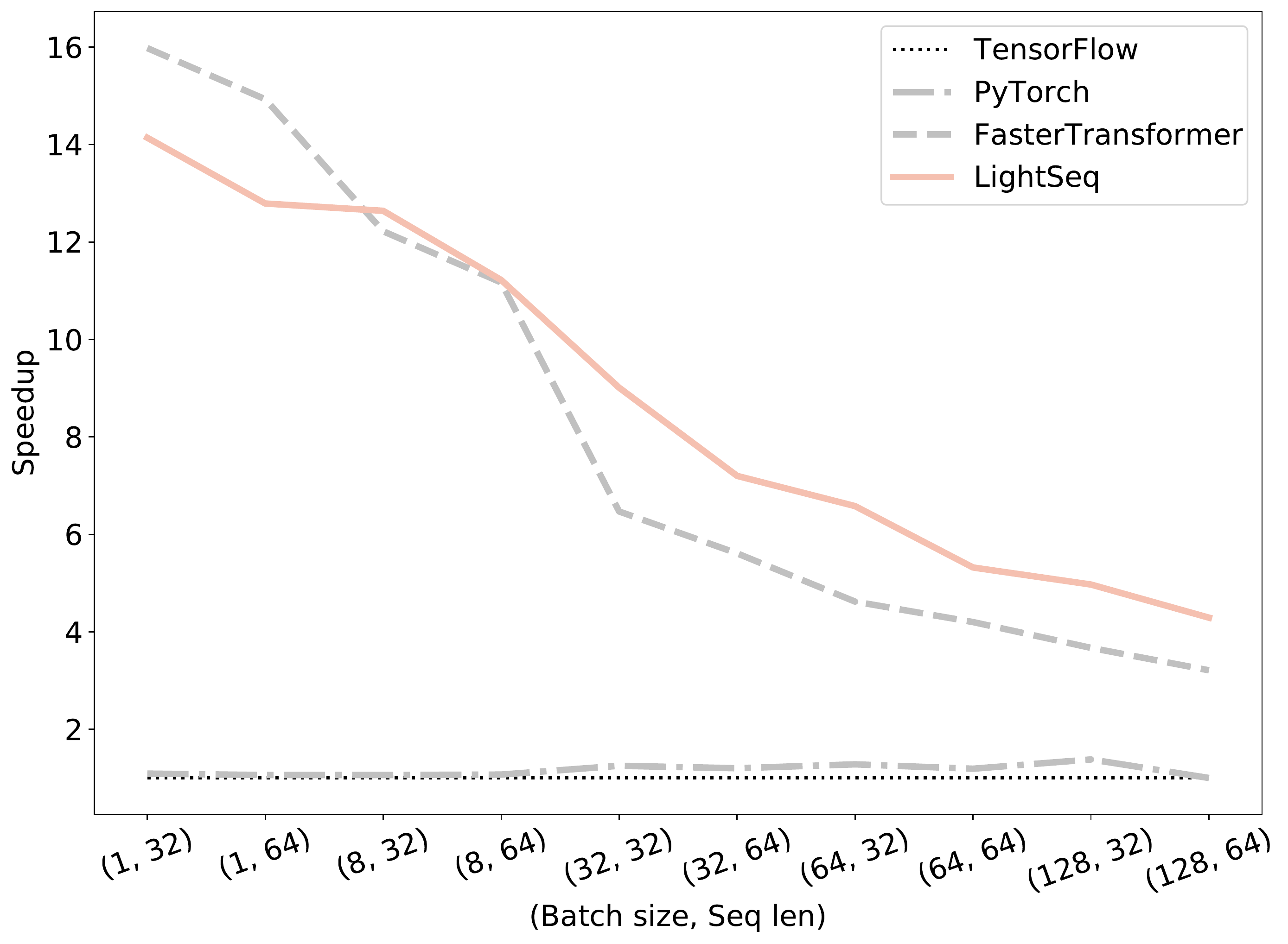}
            \end{minipage}
        }
        \caption{Speedup on Transformer with beam search compared with FasterTransformer, TurboTransformers and PyTorch implementation. The baseline is TensorFlow implementation.}
        \label{Fig:MT}
    \end{figure*}
    
    \subsection{Dynamic GPU Memory Reuse}
    In order to save GPU memory occupancy and
    avoid allocation of GPU memory during the model serving,
    \method  pre-defines the maximum of dynamic shapes, such as the maximal sequence length.
    At the start of the service, each intermediate result in the calculation process
    is allocated GPU memory to its maximum.
    Besides, GPU memory is shared for non-dependent intermediate results.

    Through this memory reuse strategy,
    on a T4 graphics card, we can deploy up to 8 Transformer big models\footnote{Under the configuration of 8 batch size, 256 sequence length, 4 beam size and 30000 vocabulary size.}
    at the same time, so as to improve graphics card utilization in low frequency or peak-shifting scenarios.

    \section{Experiments}
    \label{sec:experiment}

    In this section, we will show the improvements of \method with different GPU hardware and precisions. 
    We first analyze the GPU occupation of \method during inference to investigate if \method can make full use of GPU resources.  
    Then, we make a fair comparison with TensorFlow, PyTorch, FasterTransformer, and TurboTransformers on machine translation and text generation to show the efficiency of \method.

     \subsection{Experiment Settings}
    We test the generation performance of \method on two latest NVIDIA inference GPU Tesla P4 and T4,
    choosing TensorFlow, PyTorch, and FasterTransformer implementations as a comparison. 
    Another related library, TurboTransformers, mainly focuses on the Transformer encoder and is not powerful enough for text generation. 
    Its speedup for sequence generation compared to TensorFlow is only about $15\%$, and it only supports Float32 on GPU.
    Therefore we do not compare with it.
    


    The experiments on machine translation are conducted on the popular WMT14 English to German translation tasks.
    The hyper-parameters setting resembles transformer base model \cite{VaswaniAttend2017}.
    Specifically, we reduce the vocabulary size of both the source language and target language to 50K symbols using the sub-word technique \cite{bojanowski2017enriching}.
    
    The experiments on text generation are conducted on a randomly initialized Transformer model and test dataset.
    Results of Tensorflow and FasterTransformer are obtained from the scripts in the source code of FasterTransformer.
    The sequence length is used for limiting the total size in the generation test, and the values for  top-\(k\) and top-\(p\) are the most selected settings in our deployments.

    \subsection{GPU Occupation  of \method}
    We first analyze the GPU occupation to verify the efficiency of \method. 
    The experiments are conducted on Tesla T4 card with the GPU profiling toolkit.
    The latency of each module is shown in Figure \ref{Fig:model} with both Float16 and Float32 precision.
    We classify the operation into three categories: GEMM, cache refreshing, and others. GEMM latency is the most important indicator, which shows the proportion of matrix calculations occupying the GPU calculation.

    After optimization, we can find that:
    \begin{itemize}
        \item GEMM operation in \method accounts for 87\% and 82\% respectively for Float16 and Float32,
        accounting for most of the inference time.
        However, in the original TensorFlow model, GEMM operations account for only 25\%.
        This shows that beam search optimization has achieved good results.
        \item Cast and other operations in TensorFlow are expensive, which launches over 80 different GPU kernels. 
        In \method, we fuse cast operations into weight loading, and other operations into more efficient implementations.
        \item The latency of cache refreshing in \method accounts for 6\% and 10\% respectively,
        which are not negligible but hard to be optimized further.
        Possible solutions include reducing the amount of cache,
        such as reducing the number of decoder layers, reducing cache precision, etc.
    \end{itemize}
    
    The results demonstrate  that \method  has been optimized to a disabling extent and greatly increases the speed of inference. 
    Another interesting finding is that Float16 is more efficient than Float32.  A possible explanation  is that Float16 occupies less memory. Therefore the cache refreshing and memory I/O operations potentially take less time. 
    
    \begin{figure}[t]
        \centering
        \subfigure[Top-$p = 0.75$.]{
            \begin{minipage}[t]{0.21\textwidth}
                \includegraphics[width=\textwidth]{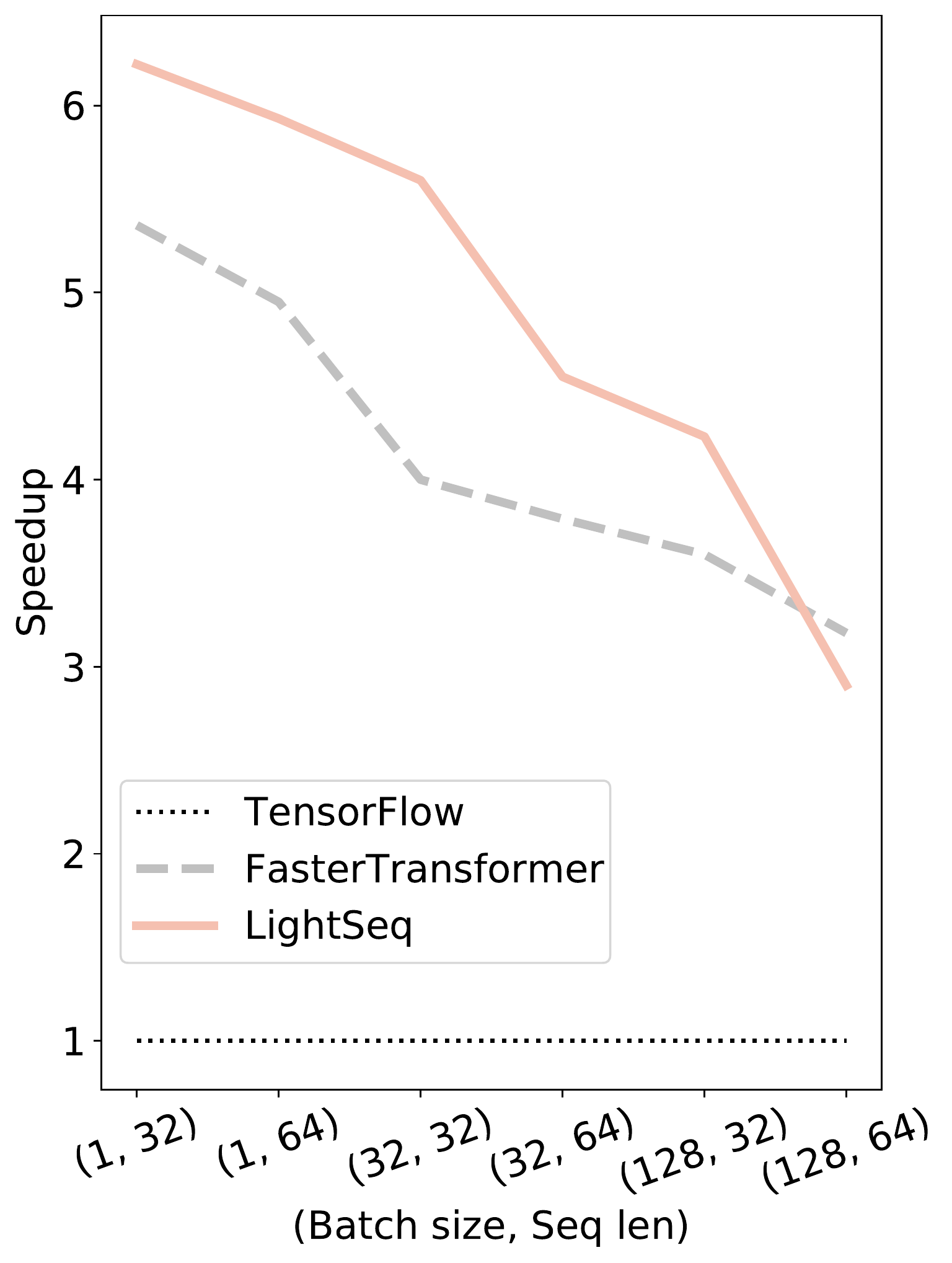}
            \end{minipage}
            \label{Fig:tg1}
        }
        \subfigure[Top-$k = 32$.]{
            \begin{minipage}[t]{0.235\textwidth}
                \includegraphics[width=\textwidth]{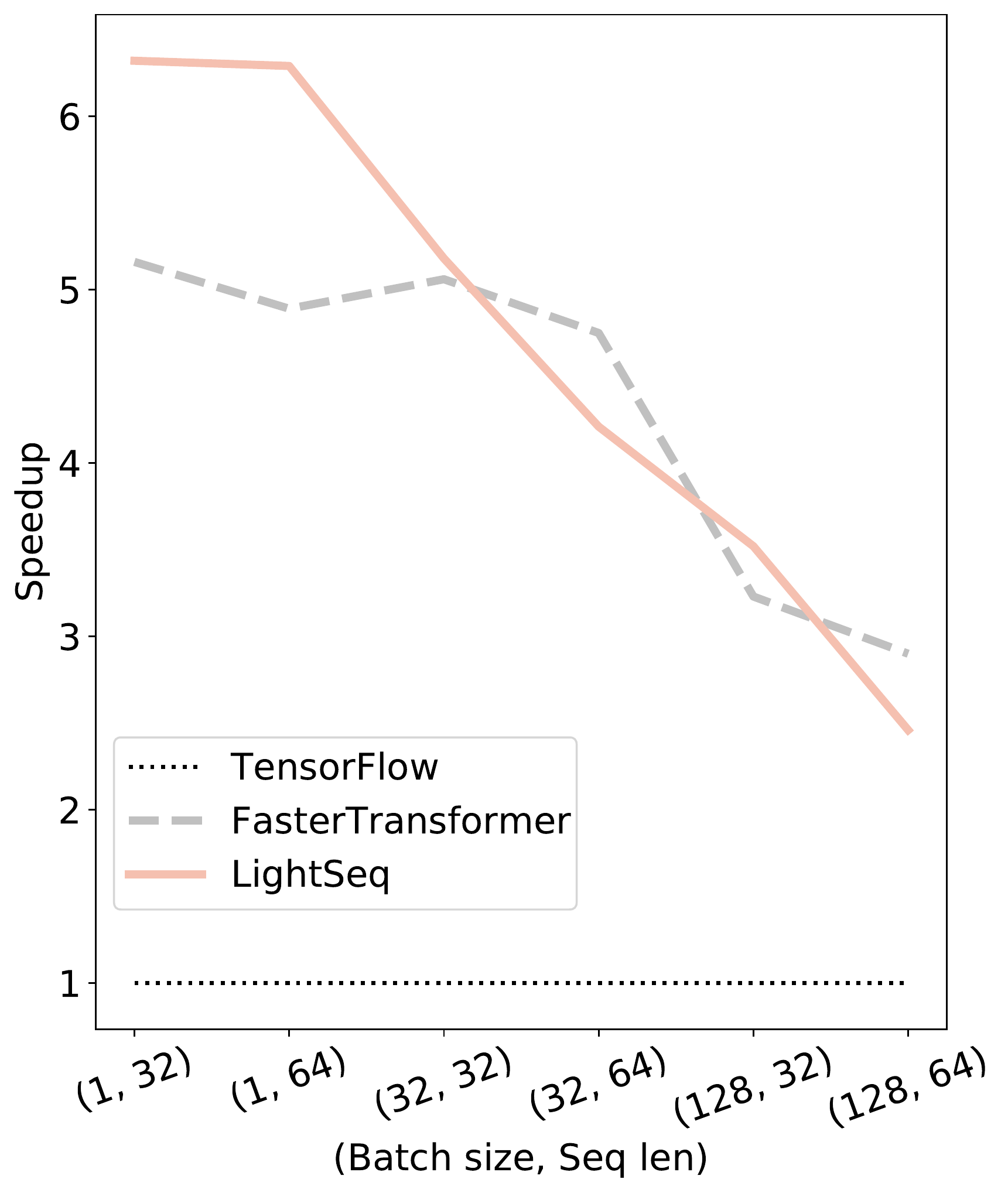}
            \end{minipage}
            \label{Fig:tg2}
        }
        \caption{T4 speedup on Transformer with sampling compared with FasterTransformer in Float16. \method outperforms FasterTransformer in most cases.}
        \label{Fig:TG}
    \end{figure}

    \subsection{Comparison  on Machine Translation}

    The comparison between \method, TensorFlow, PyTorch and FasterTransformer
    are shown in Figure  \ref{Fig:MT}.   We group the test set into different buckets according to the sequence length and batch size. 
    For example, the $x$-axis $(a,b)$ indicates that the batch size is $a$ and the sequence length  is $b$. 
    The $y$-axis is the speedup compared with TensorFlow baseline.
    The results provide several interesting findings:
    \begin{itemize}

        \item For both \method and FasterTransformer, the speedup gap for smaller batch size or shorter sequence length is much larger. 
        \item The speedup for T4 is larger than P4. The main reason is that T4 is more powerful than P4 and has much room for improvement. 
        \item In most cases, \method performs better than FasterTransformer.  For larger  batch size and longer sequences, the gap increases.  While for smaller batch size, FasterTransformer performs better. 
        \item PyTorch is slightly slower than TensorFlow in P4 and faster in T4, which indicates that LightSeq also greatly outperforms PyTorch in all cases.
    \end{itemize}
    
    The findings provide some guidance for optimization work in the future.
    There is almost no space to accelerate the inference by fusion of non-computationally intensive operators, especially for small batch size.
    Future work is recommended to focus on optimizing GEMM operations which account for 80\% to 90\% of the total computation time.

    Finally, we compare TurboTransformers with PyTorch by the translation demo\footnote{\url{https://github.com/TurboNLP/Translate-Demo/tree/443e6a46fefbdf64282842b6233a8bd0a22d6aeb}}. As of this writing, only decoder layers of MT Transformer in float32 precision is supported, so we only compare the latencies of decoder layers without beam search and cache refreshing. In the final results, TurboTransformers only achieves about 2x speedup for different batch sizes and sequence lengths. So TurboTransformers has no comparability with LightSeq in machine translation tasks (As TurboTransformer repo says, ``TurboTransformer will bring 15.9\% performance improvements on RTX 2060 GPU. We are still working on decoder model optimization.'').
    

 \subsection{Comparison on Text Generation}
    In the text generation scenario,   
    the sampling strategy is applied to improve the diversity of generation. Among which, top-\(k\) and top-\(p\) sampling strategies are more popular. 
    
    Figure \ref{Fig:TG} shows the performance comparison of
    Transformer base with top-\(k\)/top-\(p\) sampling. The values of top-\(k\) and top-\(p\) are added in the $x$-axis. The results provide following findings:
    \begin{itemize}

        \item In most cases, \method achieves greater speedup than FasterTransformer. Unlike results in machine translation, \method performs better for smaller batch size and shorter sequence, while FasterTransformer performs better for larger batch size and longer sequence.
        \item The speedup in generation tasks are not as large as machine translation. It is mainly because of the lower complexity of sampling methods than beam search, reducing the benefits obtained from operation fusion and HARS.
    \end{itemize}



    


    \section{Conclusion}
    \label{sec:conclusion}
    In this paper, we address the deployment problem of expensive sequence models and present an efficient inference library \method for sequence processing and generation, reducing the gap between the performance of big models and the requirement of online services. Comparisons with \fastertransformer show that we perform better in both machine translation and text generation.
In future work, we will focus on exploring more techniques to achieve a more significant speedup, including efficient integer-arithmetic-only inference and sparse GEMM computations.
    
    \section*{Acknowledgments} We would like to thank the colleagues in machine translation service and advertisement service to support our experiments in online environments and apply \method into real-time systems.

\bibliography{naacl2021,custom}
\bibliographystyle{acl_natbib}

\end{document}